\documentclass{optica-article}

\journal{opticajournal} 

\articletype{Research Article}

\usepackage{lineno}
\usepackage{graphicx}
\usepackage{cite}
\usepackage{hyperref}
\usepackage{color}
\usepackage{amsmath}
\usepackage{here}
\usepackage{epstopdf}
\newcommand{\be}{\begin{equation}}
\newcommand{\ee}{\end{equation}}
\newcommand{\bea}{\begin{eqnarray}}
\newcommand{\eea}{\end{eqnarray}}

\hypersetup
{
    pdfauthor={John Hancock},
    pdfsubject={Some subject},
    pdftitle={Sample document},
    pdfkeywords={LaTeX, PDF, hyperlinks}
}

\UseRawInputEncoding 

\begin{document}

%

\title{Multi-Mode Optical Chirality Extremizations on Incident Momentum Sphere}

\author{Pengxiang Wang,\authormark{1} Yuntian Chen,\authormark{1,$\dag$} and Wei Liu\authormark{2,3,*}}

\address{\authormark{1}School of Optical and Electronic Information, Huazhong University of Science and Technology, Wuhan, Hubei 430074, People's Republic of China\\
\authormark{2}College for Advanced Interdisciplinary Studies, National University of Defense Technology, Changsha, Hunan 410073, P. R. China.\\
\authormark{3}Nanhu Laser Laboratory and Hunan Provincial Key Laboratory of Novel Nano-Optoelectronic Information Materials and
Devices, National University of Defense Technology, Changsha, Hunan 410073, P. R. China.}
\email{\authormark{$\dag$}yuntian@hust.edu.cn}
\email{\authormark{*}wei.liu.pku@gmail.com} 

\begin{abstract*}
We study the momentum-space evolutions for chiral optical responses of multi-mode resonators scattering plane waves of varying incident directions. It was revealed, in our previous study [Phys. Rev. Lett. \textbf{126}, 253901 (2021)], that for single-mode resonators the scattering optical chiralities characterized by circular dichroism ($\mathbf{CD}$) are solely decided by the third Stokes parameter distributions of the quasi-normal mode (QNM) radiations:  $\mathbf{CD}=\mathbf{S}_3$. Here we extend the investigations to multi-mode resonators, and explore numerically the dependence of optical chiralities on incident directions from the perspectives of QNM radiations and their circular polarization singularities.  In contrast to the single-mode regime,  for multi-mode resonators it is discovered that $\mathbf{CD}$s defined in terms of extinction, scattering and absorption generally are different and cannot reach the ideal values of $\pm 1$ throughout the momentum sphere. Though the exact correspondence between $\mathbf{CD}$ and $\mathbf{S}_3$ does not hold anymore in the multi-mode regime, we demonstrate that the positions of the polarization singularities still serve as an efficient guide for identifying those incident directions where the optical chiralities can be extremized. 
\end{abstract*}


\section{Introduction}
Optical and geometric chiralities are profoundly different concepts: the former generally corresponds to distinct optical responses of photonic structures that are excited by sources of opposite handedness; while the latter is a pure mathematical (geometric) one, which indicates that an object is not invariant under the mirror operation~\cite{BARRON_2009__Molecular,BORISKINA__Singular}. Nevertheless, those concepts have almost always been discussed together, signifying the widely-spread dogma that they are inextricably linked. Chiral optics dates back partly to Louis Pasteur's investigations into chiral molecules, which are randomly orientated in the solution and thus the optical chiralities observed were effectively orientation-averaged and thus independent on the incident direction of the source~\cite{BARRON_2009__Molecular}. For such systems containing many randomly-orientated molecules, the absence of geometric chirality of each consisting molecule inevitably results in the absence the optical chirality, which is required by the fundamental law of parity conservation of Maxwell equations~\cite{BARRON_2009__Molecular,LEE_Phys.Rev._Questiona}. While for other systems,  either individual structures or ensembles of ordered or partially-ordered consisting molecules, the connection between optical and geometric chiralities are generally broken: (extrinsic) optical chiralities can be observed for achiral structures~\cite{BORISKINA__Singular,PAPAKOSTAS_Phys.Rev.Lett._optical_2003,PLUM_Phys.Rev.Lett._metamaterials_2009,CHEN_Phys.Rev.Lett._Extremize} and a chiral structure of fixed geometric chirality can manifest no optical chirality or even optical chiralities of opposite handedness for different incident directions~\cite{CHEN_Phys.Rev.Lett._Extremize}.

Optical chiralities are generally dependent on the incident directions of exciting sources, and all those directions constitute a closed momentum sphere. According to the extreme value theorem~\cite{RUDIN_1976__Principles}, irrespective of the specific physical configurations, there must be directions along which the optical chirality reaches its extremized (maximum or minimum) values.  In the single-QNM regime, that is when there is only one QNM effectively excited,  the optical chirality is extremized to its ideal values along the directions
where the polarization singularities of the QNM radiations locate  (independent on the geometric chirality of the photonic structures investigated) $\mathbf{CD}=\mathbf{S}_3$~\cite{CHEN_Phys.Rev.Lett._Extremize}: for QNM radiation directions of circular polarizations ($\mathbf{S}_3=\pm 1$), the \textbf{CD} defined ($\mathbf{CD} \in [-1,1]$) reaches its ideal values of  $\mathbf{CD}=\pm 1$; while for directions of linear polarizations ($\mathbf{S}_3=0$), optical chiralities are absent ($\mathbf{CD}= 0$). It is worth mentioning that though the principles revealed in Ref.~\cite{CHEN_Phys.Rev.Lett._Extremize} and the exact correspondence $\mathbf{CD}=\mathbf{S}_3$  are secured by the fundamental laws of electromagnetic reciprocity and energy conservation (optical theorem),  they are in principle only applicable in the single-QNM regime.

In this paper we study numerically momentum-space chirality extremizations for scattering photonic structures that support simultaneously multi-QNMs, including both  non-degenerate QNMs that are spectrally close and degenerated QNMs that are induced by geometric symmetry. The optical chiralities are characterized by \textbf{CD}s defined in terms of distinctions between extinction, scattering and absorption cross sections with incident circularly-polarized plane waves of opposite handedness (\textbf{Section \ref{single-mode}}).  Similar to the single-QNM regime, on the closed momentum space,  there must be directions along which the \textbf{CD} defined can reach their extremized values $\mathbf{CD}^{\rm{max,min}}$:  $\mathbf{CD}^{\rm{min}} \leq\mathbf{CD}\leq\mathbf{CD}^{\rm{max}}$. Nevertheless, in contrast to the single-QNM scenario, in the multi-QNM regime (\textbf{Section \ref{multi-mode}}): $\mathbf{CD}$ generally cannot reach its ideal values of $\mathbf{CD}=\pm 1$; \textbf{CD}s defined in terms of extinction, scattering and absorption are generally not identical; the correspondence between $\mathbf{CD}$ and $\mathbf{S}_3$   ($\mathbf{CD}=\mathbf{S}_3$) does not hold anymore, implying that $\mathbf{CD}$ is extremized along the directions where the polarizations of the QNM radiations are not necessarily singular. Despite those differences between single- and multi-QNM excitations,  we have demonstrated numerically that in the multi-QNM regime, the positions of the radiation polarization singularities in the momentum sphere still provide an efficient guide for identifying the directions incident along which the  $\mathbf{CD}$s can be extremized.  Our study can stimulate further explorations to establish a comprehensive QNM-based theoretical model that can systematically deal with not only reciprocal but also non-reciprocal multi-mode chirality extremizations, with incident waves structured both spatially and temporally beyond the plane-wave excitation regime.

\section{Definitions and Single-Mode Optical Chirality Extremizations}
\label{single-mode}

\begin{figure}[tp]
\centerline{\includegraphics[width=12cm]{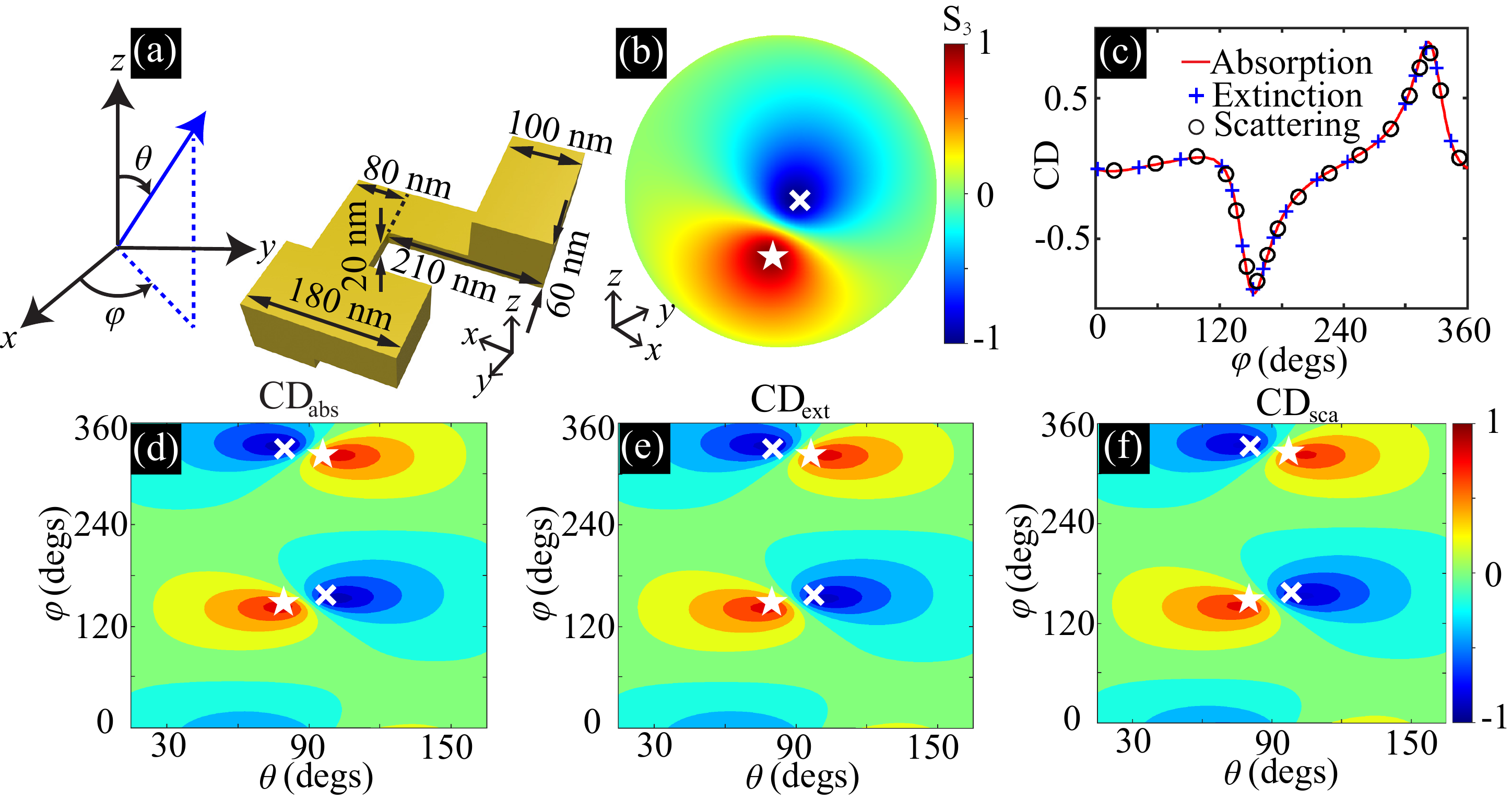}} \caption{\small (a) A gold structure [half of the four-fold rotionally symmetric structure shown in Fig.~\ref{fig4}(a)] with all geometric paramters and its orientation within the spherical coordinate system specified. The azimuthal and polar angles  are $\varphi$ and $\theta$, respectively. (b) $\mathbf{S}_3$ distributions for the QNM supported.  (c)  The dependence of three $\mathbf{CD}$s on $\varphi$ along the latituide circle of $\theta=100^\circ$. (d)-(f) The dependence of three $\mathbf{CD}$s on incident directions. In (b) and (d)-(f) the positions of polarization singularities are also marked (stars: RCP; crosses: LCP). }
\label{fig1}
\end{figure}

Throughout this study, we focus on scattering  structures, the optical responses of which are characterized by cross sections of extinction, scattering and absorption are $\rm {C}_{\rm{ext,sca,abs}}$~\cite{Bohren1983_book}.  For  right- and left-handed circularly polarized  incident plane waves (RCP and LCP, denoted respectively by $\circlearrowright$ and $\circlearrowleft$) with wavevector $\mathbf{k}_{\rm{inc}}$, those cross sections are denoted more specifically as
$\rm {C}_{\rm{ext,sca,abs}}^{\circlearrowright,\circlearrowleft}$. Then the three \textbf{CD}s defined in terms of extinction, scattering and absorption are~\cite{CHEN_Phys.Rev.Lett._Extremize}:
\begin{equation}
\label{CD-original}
\rm{\mathbf{CD}}_{\rm{ext,sca,abs}}=\frac{\rm {C}_{\rm{ext,sca,abs}}^{\circlearrowright}-\rm {C}_{\rm{ext,sca,abs}}^{\circlearrowleft}}{\rm {C}_{\rm{ext,sca,abs}}^{\circlearrowright}+\rm {C}_{\rm{ext,sca,abs}}^{\circlearrowleft}},
\end{equation}
which is bounded as $\mathbf{CD}_{\rm{ext,sca,abs}} \in [-1,1]$.

In the single-QNM excitation regime, it is revealed the three \textbf{CD}s are identical $\mathbf{CD}_{\rm{ext,sca,abs}}=\mathbf{CD}$ and inextricably linked to the Stokes parameter $\mathbf{S}_3$ of the QNM radiation (radiation wavevector $\mathbf{k}_{\rm{rad}}$) along both directions parallel or antiparallel to $\mathbf{k}_{\rm{inc}}$:
\begin{equation}
\label{CD}
\rm{\mathbf{CD}}=\mathbf{S}_3,~~\mathbf{k}_{\rm{rad}}=\pm \mathbf{k}_{\rm{inc}}.
\end{equation}
That is, in the single-QNM regime, the $\mathbf{S}_3$ for radiations along both directions of $\pm\mathbf{k}_{\rm{rad}}$ are the same.  As a result, the $\mathbf{CD}$ reaches its ideal extremized values of $\pm1$ along directions where the QNM radiations are circularly-polarized (polarization singularities)~\cite{CHEN_Phys.Rev.Lett._Extremize}. It is worth mentioning that beyond the single-QNM regime, such defined three \textbf{CD}s do not have to be identical and could be quite different. For example, for an arbitrary inversion-symmetric structure that is simultaneously reciprocal, along any incident direction  $\mathbf{CD}_{\rm{ext}}=0$ while generally $\mathbf{CD}_{\rm{sca,abs}}\neq0$~\cite{CHEN_2020_Phys.Rev.Research_Scatteringa}.

To further exemplify the conclusions above, we study a gold  structure (effective bulk permittivity is fitted from data in Ref.~\cite{Johnson1972_PRB}; all numerical results presented in this study are obtained using COMSOL Multiphysics) schematically shown in Fig.~\ref{fig1}(a), with all the geometric parameters specified.   This structure is essentially half of the structure shown in Fig.~\ref{fig4}(a) which exhibits rotation symmetry and thus supports degenerate modes~\cite{HOPKINS_2016_LaserPhotonicsRev._Circular}.  With the rotation symmetry broken,  for this structure a single QNM [characterized by a complex eigenfrequency of $\tilde{\omega}_{A}=(8.196\times10^{14}-3.5\times10^{13}\rm{i})$ rad/s] can be excited with negligible excitations of other QNMs.  $\mathbf{S}_3$ distributions for radiations of this QNM are shown in Fig.~\ref{fig1}(b), where the positions of circular polarization singularities are also marked (stars: RCP; crosses: LCP). To show clearly the orientation of the structure, in Fig.~\ref{fig1}(a) a spherical coordinate system with the azimuthal angle $\varphi$ and polar angle $\theta$ is also included. 

For all the following discussions, the incident direction  $\mathbf{k}_{\rm{inc}}$ is described by the angle doublet ($\varphi$,  $\theta$) and the corresponding radiation direction is anti-parallel to it $\mathbf{k}_{\rm{rad}}=-\mathbf{k}_{\rm{inc}}$. Such a setting is required to directly apply the electromagnetic reciprocity principle~\cite{LANDAU_1984__Electrodynamicsb,POTTON_Rep.Prog.Phys._reciprocity_2004}, built on which the subtle connection between \textbf{CD} and $\mathbf{S_3}$ of QNM radiations [Eq.~(\ref{CD})] can be established. The dependence of \textbf{CD}s on ($\varphi$,  $\theta$) are shown in Figs.~\ref{fig1}(d)-(f), with incident angular frequency $\omega=\mathbf{Re}(\tilde{\omega}_{A})$ and the corresponding wavelength $\lambda_A=2.3~\mu$m. As we can see, all three \textbf{CD}  are identical and they reach their ideal values of $\mathbf{CD}=\pm 1$ at the positions where the QNM radiations are singularly polarized [marked by stars (RCP; $\mathbf{CD=S_3}=1$) and crosses (LCP; $\mathbf{CD=S_3}=-1$)].  As has been mentioned already, the angular positions for the radiation singularities are 
 ($\pi+\varphi$,  $\pi-\theta$) rather than ($\varphi$,  $\theta$) directly. The \textbf{CD} distributions on a latitude circle ($\theta=100^\circ$) are also shown in Fig.~\ref{fig1}(c), further confirming Eq.~(\ref{CD}) and the equality of all \textbf{CD}s.
 
\begin{figure}[tp]
\centerline{\includegraphics[width=12cm]{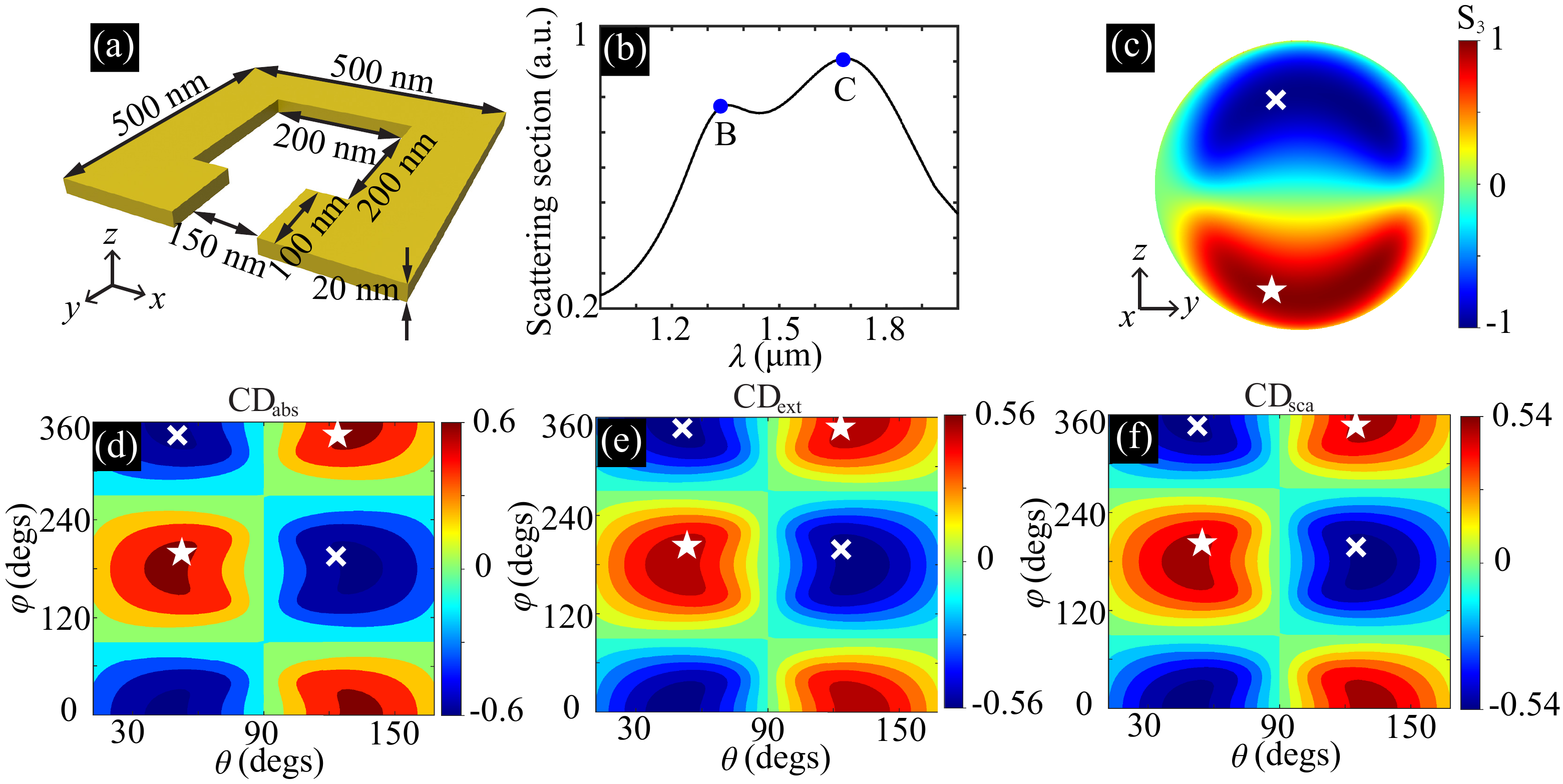}} \caption{\small (a) A gold SRR with all geometric parameters specified and (b) its scattering spectra where two QNMs are pinpointed. 
 (c) $\mathbf{S}_3$  distributions for the QNM-B. (d)-(f) The angular dependence of three $\mathbf{CD}$s. In (c)-(f) the positions of polarization singularities are also marked (stars: RCP; crosses: LCP).}
\label{fig2}
\end{figure}

\section{Multi-Mode Optical Chirality Extremizations}
\label{multi-mode}

As a next step, we extend our investigations to photonic structures that support spectrally close non-degenerate modes and also degenerate modes. We start with a gold split-ring resonator (SRR) schematically shown in Fig.~\ref{fig2}(a), with all geometric parameters specified. In Fig.~\ref{fig2}(b) we show its scattering spectra (in terms of scattering cross sections) for an RCP wave incident along \textbf{+z} axis ($\theta=0$). For the spectral regime shown, two QNMs are marked the corresponding complex eigenfrequencies are:  $\tilde{\omega}_{B}=(1.4359\times10^{15}-7.7\times10^{13}\rm{i})$ rad/s and  $\tilde{\omega}_{C}=(1.1204\times10^{15}-2.538\times10^{14}\rm{i})$ rad/s. The $\mathbf{S}_3$ distributions for the QNM-B are shown in Fig.~\ref{fig2}(c). We now fix the angular frequency of the incident wave as $\omega=\mathbf{Re}(\tilde{\omega}_{B})$ ($\lambda_B=1.313~\mu$m) and show  in Figs.~\ref{fig2}(d)-(f) the angular dependence of \textbf{CD}s, where the polarization singularities of QNM-B also marked. In contrast to the single-QNM excitations [Figs.~\ref{fig1}(d)-(f)]: (i) The three \textbf{CD}s are not identical anymore (see the extreme values of the color bar); (ii) None of the \textbf{CD}s defined can reach their ideal values of $\pm 1$; (iii) The positions of the extremized \textbf{CD}s ($\mathbf{CD}^{\rm{min}}$ and $\mathbf{CD}^{\rm{max}}$) do not overlap exactly at those of the polarization singularities. Despite those discrepancies, it is clear from Figs.~\ref{fig2}(d)-(f) that the positions of the QNM singularities still provide an excellent guide for identifying the directions along which the extremized values of all \textbf{CD}s can be obtained.

\begin{figure}[tp]
\centerline{\includegraphics[width=12cm]{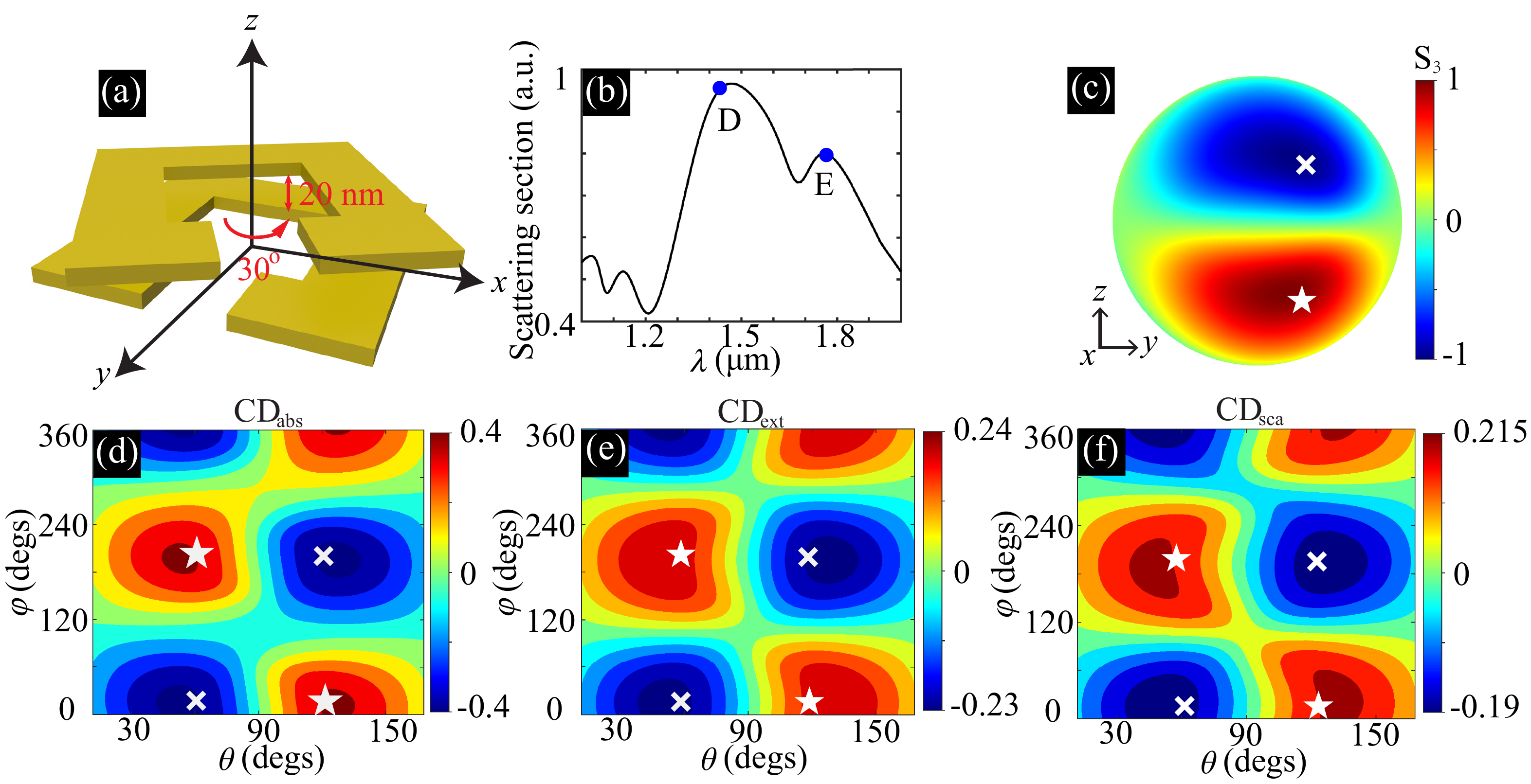}} \caption{\small (a) A gold SRR-pair with a twisting angle $30^\circ$  [each SRR is idential to that shown in Fig.~\ref{fig2}(a)] and (b) its scattering spectra where two QNMs are pinpointed. (c) $\mathbf{S}_3$  distributions for the QNM-D. (d)-(f) The angular dependence of three $\mathbf{CD}$s. In (c)-(f) the positions of polarization singularities are also marked (stars: RCP; crosses: LCP).}
\label{fig3}
\end{figure}

We now proceed to the more sophisticated configuration of a coupled SRR pair ( twisted by $30^\circ$ with respect to each other) shown in Fig.~\ref{fig3}(a), where each SRR is identical to the one shown in Fig.~\ref{fig3}(a). Its scattering cross section spectra (RCP wave incident along \textbf{+z} axis) are shown in Fig.~\ref{fig3}(b), where two spectrally close QNMs are marked. Their corresponding complex eigenfrequencies are:  $\tilde{\omega}_{D}=(1.357\times10^{15}-2\times10^{14}\rm{i})$ rad/s and $\tilde{\omega}_{E}=(1.0279\times10^{15}-3.23\times10^{14}\rm{i})$ rad/s. The $\mathbf{S}_3$ distributions for the QNM-D are shown in Fig.~\ref{fig3}(c). Figures~\ref{fig3}(d)-(f) demonstrate the angular dependence of \textbf{CD}s with $\omega=\mathbf{Re}(\tilde{\omega}_{D})$ ($\lambda_D=1.39~\mu$m), where the positions of the polarization singularities for QNM-D are also marked. Similar to the single SRR scenario, though the positions of the polarization singularities do not overlap exactly with those of extremized \textbf{CD}s, they still served as an efficient guide to identify the optimal incident directions for chirality extremizations.

\begin{figure}[tp]
\centerline{\includegraphics[width=12cm]{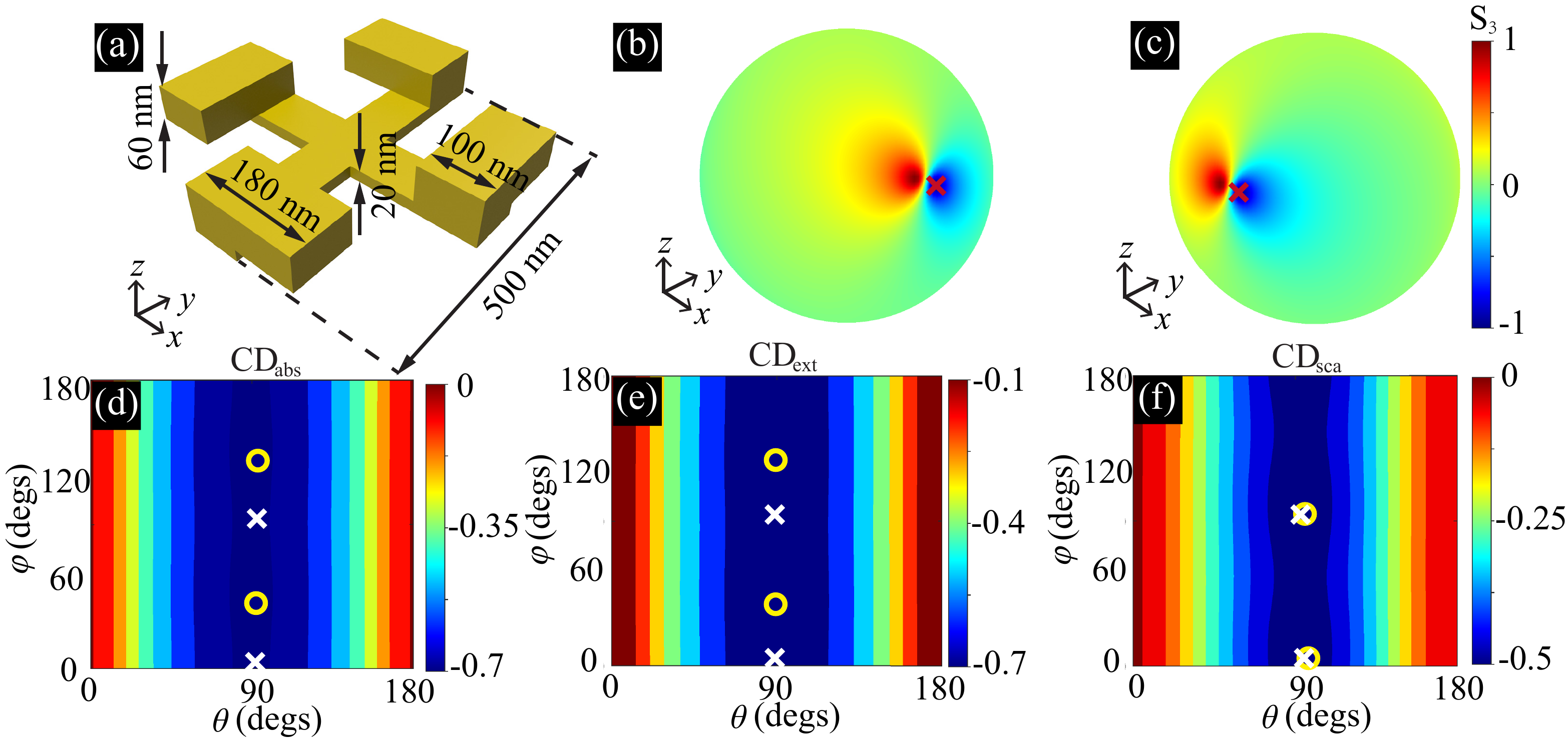}} \caption{\small (a) A gold structure exhibiting four-fold rotation symmetry, with all geometric parameters specified. (b)-(c) $\mathbf{S}_3$ distributions for the two degenerate QNMs. (d)-(f) The angular dependence of three $\mathbf{CD}$s. In (b)-(f) the positions of LCP singularities are also marked by crosses. In (d)-(f) the actual positions of the minimum values of $\mathbf{CD}$ ($\mathbf{CD}^{\rm{min}}$) are also marked by circles.}
\label{fig4}
\end{figure}

At the end, we turn to the gold structure [shown in Fig.~\ref{fig4}(a)] that exhibits four-fold rotation symmetry. Such a symmetry secures a pair of degenerate modes~\cite{HOPKINS_2016_LaserPhotonicsRev._Circular}. The $\mathbf{S}_3$ distributions of both are shown in Figs.~\ref{fig4}(b) and (c) and their degenerate complex eigenfrequency is  $\tilde{\omega}_{F}=(7.9645\times10^{14}-4.17\times10^{13}\rm{i})$ rad/s. With the angular frequency of the incident wave fixed as $\omega=\mathbf{Re}(\tilde{\omega}_{F})$ ($\lambda_F=2.367~\mu$m), the angular dependence of \textbf{CD}s is summarized in Figs.~\ref{fig4}(d)-(f), with the LCP singularities of both QNMs marked. Since all \textbf{CD}s are negative, half of the polarization singularities (RCP; $\mathbf{S_3}=1$) have totally lost their connections with $\mathbf{CD}^{\rm{max}}$.  This is probably due to the co-excitation of both QNMs and that none of them plays a overwhelming role, while a precise clarification for it, from the perspectives of QNMs with explicit formalism, is yet to be explored. Despite this, the positions of LCP singularities still provide reasonable guide for identifying the optimal directions to obtain $\mathbf{CD}^{\rm{min}}$ [positions marked by circles in Figs.~\ref{fig4}(d)-(f)], especially efficient for $\mathbf{CD}_{\rm{sca}}$ shown in Fig.~\ref{fig4}(f).

\section{Conclusions}
In this work, we extend our pervious investigations of single-mode chirality extremizations to the multi-mode regime, studying scattering structures when more than one QNMs are simultaneously excited. The optical chiralities are characterized by \textbf{CD}s defined in terms of extinction, scattering and absorption, which are not identical anymore beyond the single-mode regime. We have further revealed that though the exact correspondence between \textbf{CD}s and $\mathbf{S}_3$ distributions of the QNM radiations is broken, the positions of the polarization singularities of the QNMs excited still provide an efficient guide to identify those directions incident along which the \textbf{CD}s can be extremized. Our numerical studies presented in this work can stimulate further in-depth explorations for a theoretical model that can systematically deal with the problem of momentum-space optical chirality extremizations, not only beyond the single-QNM regime, but also covering spatially and spectrally structured incident waves (\textit{e.g.} those carrying both spin and angular momentum~\cite{CHEN_2022_NatRevPhys_Multidimensional}), for scattering photonic structures that are not necessarily reciprocal.

\section*{Funding}
National Natural Science Foundation of China (Grant No. 11404403 and 61405067); National Key Research and Development Program of China (Grant No. 2021YFB2800303);Outstanding Young Researcher Scheme of National University of Defense Technology of China; Innovation Project of Optics Valley Laboratory.
\section*{Acknowledgments}
We thank Weijin Chen and  Qingdong Yang for their help with numerical simulations.

\bibliographystyle{opticajnl}
\bibliography{References_scattering_4}

\begin{thebibliography}{10}
\newcommand{\enquote}[1]{``#1''}

\bibitem{BARRON_2009__Molecular}
L.~D. Barron, \emph{Molecular {{Light Scattering}} and {{Optical Activity}}}
  ({Cambridge University Press}, 2009).

\bibitem{BORISKINA__Singular}
S.~Boriskina and N.~I. Zheludev, eds., \emph{Singular and {{Chiral
  Nanoplasmonics}}} ({Jenny Stanford Publishing}, {Singapore}, 2014), 1st ed.

\bibitem{LEE_Phys.Rev._Questiona}
T.-D. Lee and C.-N. Yang, \enquote{Question of parity conservation in weak
  interactions,} {\protect\JournalTitle{Phys. Rev.}} \textbf{104}, 254 (1956).

\bibitem{PAPAKOSTAS_Phys.Rev.Lett._optical_2003}
A.~Papakostas, A.~Potts, D.~M. Bagnall, S.~L. Prosvirnin, H.~J. Coles, and
  N.~I. Zheludev, \enquote{Optical manifestations of planar chirality,}
  {\protect\JournalTitle{Phys. Rev. Lett.}} \textbf{90}, 107404 (2003).

\bibitem{PLUM_Phys.Rev.Lett._metamaterials_2009}
E.~Plum, X.-X. Liu, V.~A. Fedotov, Y.~Chen, D.~P. Tsai, and N.~I. Zheludev,
  \enquote{Metamaterials: Optical activity without chirality,}
  {\protect\JournalTitle{Phys. Rev. Lett.}} \textbf{102}, 113902 (2009).

\bibitem{CHEN_Phys.Rev.Lett._Extremize}
W.~Chen, Q.~Yang, Y.~Chen, and W.~Liu, \enquote{Extremize {{Optical
  Chiralities}} through {{Polarization Singularities}},}
  {\protect\JournalTitle{Phys. Rev. Lett.}} \textbf{126}, 253901 (2021).

\bibitem{RUDIN_1976__Principles}
R.~W. Rudin, \emph{Principles of {{Mathematical Analysis}}} ({McGraw-Hill
  Publishing Company}, {Auckland}, 1976), 3rd ed.

\bibitem{Bohren1983_book}
C.~F. Bohren and D.~R. Huffman, \emph{Absorption and Scattering of Light by
  Small Particles} (Wiley, 1983).

\bibitem{CHEN_2020_Phys.Rev.Research_Scatteringa}
W.~Chen, Q.~Yang, Y.~Chen, and W.~Liu, \enquote{Scattering activities bounded
  by reciprocity and parity conservation,} {\protect\JournalTitle{Phys. Rev.
  Research}} \textbf{2}, 013277 (2020).

\bibitem{Johnson1972_PRB}
P.~B. Johnson and R.~W. Christy, \enquote{Optical constants of the noble
  metals,} {\protect\JournalTitle{Phys. Rev. B}} \textbf{6}, 4370 (1972).

\bibitem{HOPKINS_2016_LaserPhotonicsRev._Circular}
B.~Hopkins, A.~N. Poddubny, A.~E. Miroshnichenko, and Y.~S. Kivshar,
  \enquote{Circular dichroism induced by {{Fano}} resonances in planar chiral
  oligomers,} {\protect\JournalTitle{Laser Photon. Rev.}} \textbf{10}, 137--146
  (2016).

\bibitem{LANDAU_1984__Electrodynamicsb}
L.~D. Landau, L.~P. Pitaevskii, and E.~M. Lifshitz, \emph{Electrodynamics of
  {{Continuous Media}}: {{Volume}} 8} ({Butterworth-Heinemann}, {Amsterdam
  u.a}, 1984), 2nd ed.

\bibitem{POTTON_Rep.Prog.Phys._reciprocity_2004}
R.~J. Potton, \enquote{Reciprocity in optics,} {\protect\JournalTitle{Rep.
  Prog. Phys.}} \textbf{67}, 717 (2004).

\bibitem{CHEN_2022_NatRevPhys_Multidimensional}
Y.~Chen, W.~Du, Q.~Zhang, O.~{\'A}valos-Ovando, J.~Wu, Q.-H. Xu, N.~Liu,
  H.~Okamoto, A.~O. Govorov, Q.~Xiong, and C.-W. Qiu, \enquote{Multidimensional
  nanoscopic chiroptics,} {\protect\JournalTitle{Nat. Rev. Phys.}} \textbf{4},
  113--124 (2022).

\end{thebibliography}

\end{document}